\documentclass[10pt,superscriptaddress,aps,pra,twocolumn,nofootinbib]{revtex4-2}
\usepackage[T1]{fontenc}
\usepackage{amsmath,amssymb,amsthm,amsfonts,mathtools,bm}
\usepackage{graphicx}
\usepackage{dcolumn}
\usepackage{tabularx,booktabs,multirow,makecell}
\usepackage[breaklinks=true,colorlinks,citecolor=blue,linkcolor=red,urlcolor=blue]{hyperref}
\usepackage{xurl}
\usepackage{braket}
\usepackage[normalem]{ulem}
\usepackage{appendix}
\usepackage{tikz}
\usepackage[linesnumbered,lined,ruled,vlined]{algorithm2e}

\begin{document}
%%
%%===================THE TITLE========================
%%
\title{Shortcuts to adiabaticity for rapid soliton compression in nonlocal media}
%%
%%===================AUTHOR LIST=======================
%
%

\author{Yingjia Li}
%\thanks{\href{https://orcid.org/0000-0002-7966-0241}{ORCID: 0000-0002-7966-0241}}
\affiliation{Department of Physics, Shanghai University, Shanghai 200444, China}
\affiliation{Department of Physical Chemistry, University of the Basque Country UPV/EHU, Apartado 644, 48080 Bilbao, Spain}

\author{Qian Kong}
%\thanks{\href{https://orcid.org/0000-0003-0243-9626}{ORCID: 0000-0003-0243-9626}}
\affiliation{Department of Physics, Shanghai University, Shanghai 200444, China}

\author{Xihua Yang}
\email{yangxh@shu.edu.cn}
%\thanks{\href{https://orcid.org/0000-0003-0243-9626}{ORCID: 0000-0003-0243-9626}}
\affiliation{Department of Physics, Shanghai University, Shanghai 200444, China}

\author{Xi Chen}
\email{xi.chen@csic.es}
%\thanks{\href{https://orcid.org/0000-0003-4221-4288}{ORCID: 0000-0003-4221-4288}}
\affiliation{Quantum Advanced Research Center (QuARC), CSIC, 28049, Madrid, Spain}
\affiliation{Instituto de Ciencia de Materiales de Madrid (ICMM), CSIC, 28049, Madrid, Spain}

\date{\today}

\begin{abstract}
We investigate shortcut-to-adiabaticity protocols for the rapid compression of optical solitons in nonlocal media. Using a variational approximation, we derive an effective Ermakov-like equation for the soliton width and employ inverse engineering to design propagation-dependent control parameters. Three control strategies are compared, based respectively on modulation of the characteristic nonlocal length, the Kerr nonlinearity, and the external parabolic confinement. We show that all three protocols enable high-fidelity compression over strongly reduced propagation distances. By comparing control smoothness and final fidelity, we find that the relative performance of the protocols in the very short-distance regime depends on the chosen criterion, revealing a trade-off between implementation smoothness and target-profile accuracy. These results establish a practical and general route to fast soliton manipulation in nonlocal media.
\end{abstract}

	\maketitle
\section{Introduction}
\label{sec:1}

In nonlocal media, the nonlinear response at a given spatial position is determined not only by the local field intensity but also by the intensity distribution over a finite surrounding region \cite{krolikowski2004modulational}. This property is commonly described by a nonlocal response function, whose form may be Gaussian, exponential, or rectangular, depending on the physical system under consideration \cite{PhysRevE.64.016612}. Nonlocality has been extensively investigated both theoretically and experimentally in a variety of settings, including nematic liquid crystals (NLCs) \cite{aleksic2012solitons, peccianti2002nonlocal, conti2004observation}, thermal optical media \cite{rubin2018nonlocal}, and Bose-Einstein condensates (BECs) with long-range interactions \cite{lahaye2009physics, gligoric2010two}.

A major consequence of nonlocality is its strong influence on the existence, stability, and dynamics of solitons. Solitons are self-trapped wave packets sustained by the balance between dispersion or diffraction and nonlinearity. In local nonlinear media, solitons are often highly sensitive to perturbations and may undergo collapse or dynamical instabilities. By contrast, nonlocality introduces a spatial averaging effect that smooths intensity fluctuations and suppresses the growth of perturbations. As a result, nonlocal media can support solitons with enhanced robustness and enlarged stability domains \cite{maucher2012stability, skupin2006stability, bang2002collapse}.

Beyond improving stability, nonlocality also gives rise to a wide range of nonlinear-wave phenomena. Depending on the degree of nonlocality and the underlying medium, nonlocal solitons may exhibit breathing dynamics, oscillatory motion, effective attraction, and the formation of bound states mediated by long-range interactions rather than direct spatial overlap \cite{alberucci2016breather, garza2019soliton, skupin2006stability, briedis2005ring, esbensen2011modulational, nikolov2004attraction}. Rich families of dark, antidark, and ring-shaped nonlocal solitons have also been identified, including in systems with competing nonlinearities \cite{esbensen2011modulational, kong2013dark, horikis2016ring}. The interplay between nonlocality and external trapping potentials further enriches the dynamics, leading to transitions between particlelike transport and wave scattering, as well as other forms of controllable nonlinear propagation \cite{koutsokostas2024particle, rasmussen2005theory}. More recently, nonlocal soliton physics has been extended to systems with higher-order propagation effects, such as fourth-order diffraction, where dark-soliton solutions, interactions, and bound states have been investigated using variational and numerical approaches \cite{Shenpre}. These features make nonlocal media a versatile platform for manipulating nonlinear excitations.

The characteristic length of the nonlocal response is an important physical parameter, since it determines the spatial range over which the nonlinear response is averaged and therefore strongly affects soliton localization and stability. In several experimentally relevant platforms, this length scale can be externally tuned. For instance, in nematic liquid crystals the degree of nonlocality can be continuously controlled through the applied bias voltage, which modifies the molecular tilt-angle distribution \cite{kivshar2003optical, hu2006nonlocality, conti2004observation}. In thermal nonlinear media such as lead glass, the effective response length can be engineered through the sample geometry and thermal boundary conditions \cite{rotschild2005solitons, minovich2007experimental}. Similarly, in dipolar Bose-Einstein condensates, the relative importance of nonlocal interactions can be enhanced by reducing the s-wave scattering length via Feshbach resonances \cite{griesmaier2005bose, lahaye2007strong}. {This tunability naturally connects nonlocal soliton dynamics with the broader framework of soliton management, where engineered system parameters have been used to compress matter-wave solitons through scattering-length management \cite{liang2005dynamics}, to control soliton interactions via longitudinally varying nonlocality and nonlinearity \cite{ye2007enhanced}, and to manipulate nonautonomous soliton transport through inhomogeneous nonlocality \cite{sun2017transport}.} These examples show that the nonlocal response length is not only a defining property of the medium, but also a potential control knob for fast soliton manipulation.

From a control perspective, an important objective is to reshape or compress a soliton while preserving its integrity. A standard strategy is adiabatic manipulation, in which system parameters such as the nonlocal response range, the nonlinear coefficient, or the external confinement are varied sufficiently slowly that the soliton continuously adapts to the instantaneous stationary state without generating significant radiation or excitations \cite{Smith:89,PhysRevLett.45.1095,PhysRevLett.71.73,PhysRevE.58.6637,Turitsyn:19}. Although adiabatic protocols are robust in principle, their practical use is limited by the long propagation distances required to maintain adiabaticity. This leads to larger device footprints, slower operation, and greater sensitivity to loss and environmental perturbations \cite{peccianti2002nonlocal}. These limitations motivate faster control strategies capable of reproducing adiabatic final states within substantially reduced evolution distances \cite{LiOE24,Hanpra206}.

Shortcuts to adiabaticity (STA) provide such a route. STA protocols are designed to drive a system to the same target state that would be reached adiabatically, but within a finite and typically much shorter evolution time or propagation distance while maintaining high fidelity \cite{guery2019shortcuts}. Originally developed in quantum control, STA methods have since found broad applications in atomic, molecular, and optical physics. Their extension to nonlinear systems is particularly appealing but also nontrivial, because nonlinear dynamics generally do not admit simple dynamical invariants \cite{li2016shortcut,Tangyoupra}. One practical way to overcome this difficulty is to combine inverse engineering with a variational approximation \cite{Borisreview,KIVSHAR1995353,perez1996low}, thereby reducing the field dynamics to effective equations for a few physically relevant collective variables \cite{perez1996low,haas2018time}. This strategy has proven useful for the control of nonlinear excitations, including matter-wave solitons, and can also be generalized to multisoliton settings \cite{wang2020controllable, al2002bright}.

In this work, we develop an inverse-engineered STA framework for the rapid compression of optical solitons in nonlocal media. Starting from a generalized nonlocal nonlinear Schr\"odinger equation, we employ a Gaussian variational ansatz to derive an effective Ermakov-like equation for the soliton width. This reduced description enables the systematic construction of finite-distance STA protocols and allows us to compare different control knobs on equal footing. We investigate three strategies based on modulation of the characteristic nonlocal length, the Kerr nonlinearity, and the external parabolic confinement. We show that all three protocols can realize high-fidelity soliton compression over strongly reduced propagation distances. Moreover, by comparing control smoothness and final infidelity, we find that modulation of the characteristic nonlocal length generally yields the smoothest control profile and performs favorably over a broad range of propagation distances, while in the very short-distance regime the relative performance depends on the chosen criterion. This reveals a trade-off between implementation smoothness and target-profile accuracy.

The remainder of this paper is organized as follows. Sec.~\ref{sec:2} introduces the generalized nonlocal nonlinear Schr\"odinger equation and derives an effective width equation within a variational approximation. Sec.~\ref{section3} presents the inverse-engineered STA protocols based on modulation of the characteristic nonlocal length, the Kerr nonlinearity, and the external confinement. {Sec.~\ref{sec4} compares the resulting control smoothness, fidelity, and shortcut efficiency, and further discusses the complete compression dynamics and the emitted radiation.} Finally, Sec.~\ref{sec:5} summarizes the main conclusions and outlines possible future directions.

\section{Model and method}
\label{sec:2}

The propagation of an optical pulse in a nonlocal medium with an additional Kerr nonlinearity and an external refractive-index modulation is governed by the generalized nonlocal nonlinear Schr\"odinger equation (NNLSE) \cite{snyder1997accessible, PhysRevE.64.016612},
\begin{equation}
\begin{aligned}
i\partial_z u
&+\frac{1}{2}\partial_x^2u
+u\!\int\!R(x-\xi)|u(\xi,z)|^2\,d\xi \\
&+\gamma|u|^2u+n(x,z)u=0.
\end{aligned}
\label{NNLS}
\end{equation}
The dimensionless variables are introduced from the corresponding physical ones, denoted by tildes, as \cite{kivshar2003optical}
\begin{equation}
u=kw_0\sqrt{n_2}\tilde{u}, \qquad x=\tilde{x}/w_0, \qquad z=\tilde{z}/z_R.
\end{equation}
Here, $\tilde{u}$ is the complex envelope of the electric field, while $\tilde{x}$ and $\tilde{z}$ denote the transverse and longitudinal coordinates, respectively. The beam waist is denoted by $w_0$, and the Rayleigh length is $z_R=kw_0^2$, with $k$ the wave number in the medium. The nonlinear index coefficient $n_2$ characterizes the medium response, with $n_2>0$ corresponding to a self-focusing medium and $n_2<0$ to a self-defocusing one. The dimensionless parameter $\gamma=\tilde{\gamma}/(kn_2)$ describes the local Kerr nonlinearity, for which the refractive-index change is proportional to $|u|^2$. The function $n(x,z)$ represents an externally imposed refractive-index landscape modeling a tunable waveguide trap, which we take here to be parabolic,
\begin{equation}
\label{parabolic}
n(x,z)=-\alpha^2 x^2,
\end{equation}
where $\alpha$ is a dimensionless parameter characterizing the waveguide strength. In principle, gain and loss terms may also be included in the model. However, after an appropriate transformation, their effect can be mapped onto an effective modification of the nonlinear response \cite{LiOE24}. For this reason, and in order to focus on the role of nonlocality and shortcut control, we restrict ourselves to the conservative case in the present work.

The nonlocal response function $R(x)$ is assumed to be real, symmetric, and normalized according to $\int_{-\infty}^{+\infty}R(x)dx=1$. For analytical convenience, and also to ensure modulational stability in self-focusing media \cite{PhysRevE.64.016612, esbensen2011modulational}, we adopt a Gaussian response function,
\begin{equation}
R(x)=\frac{1}{\sqrt{\pi}\sigma(z)}\exp\left[-\frac{x^2}{\sigma^2(z)}\right].
\end{equation}
Here, $\sigma$ denotes the characteristic length scale of the nonlocal response. Physically, it determines the spatial range over which the nonlinear response is averaged. In the limit $\sigma\to 0$, Eq.~(\ref{NNLS}) reduces to the nonlinear Schr\"odinger equation for a local medium.

\begin{figure}
    \centering
    \includegraphics[width=1.\linewidth]{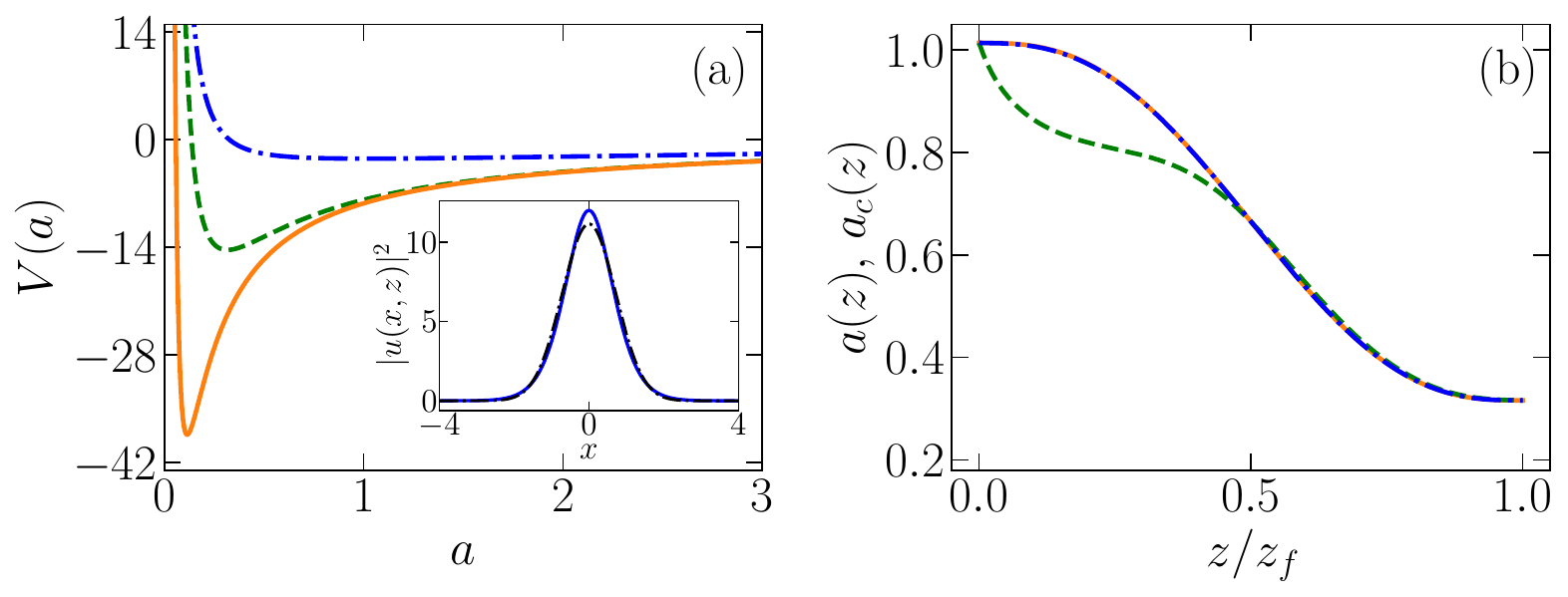}
    \caption{(a) Effective potential $V(a)$ as a function of the width parameter $a$ for three different values of the nonlocal length: $\sigma=5$ (blue dash-dotted), $\sigma=0.5$ (green dashed), and $\sigma=0.01$ (orange solid). The inset compares the stationary initial soliton obtained by imaginary-time propagation (blue solid) with the initial Gaussian ansatz in Eq.~(\ref{eq:gaussian_ansatz}) (black dash-dotted). (b) Comparison between the prescribed minimum-jerk compression trajectory $a(z)$ and the corresponding reference width $a_c(z)$. For $z_f=50$, the prescribed trajectory (orange solid) and the reconstructed reference width obtained from the designed $\sigma(z)$ through Eq.~(\ref{ermakov}) (blue dash-dotted) are nearly indistinguishable, indicating adiabatic evolution. For $z_f=3$, the corresponding reference width $a_c(z)$ deviates significantly from the prescribed trajectory, reflecting the breakdown of adiabatic following. The parameters used here are $a_i=1.014$, $a_f=0.316$, $\gamma=0.1$, $\alpha=0.1$, and $P=20$.}
    \label{effective_potential}
\end{figure}

To investigate the soliton dynamics analytically for an arbitrary degree of nonlocality, we adopt the Gaussian variational ansatz
\begin{equation}
u(x,z)=A(z)\exp\left[-\frac{x^2}{2a^2(z)}+ib(z)x^2+i\phi(z)\right],
\label{eq:gaussian_ansatz}
\end{equation}
where $A(z)$, $a(z)$, $b(z)$, and $\phi(z)$ represent the amplitude, width, chirp, and phase of the beam, respectively. In what follows, we omit the explicit $z$ dependence whenever no confusion arises, e.g., $a(z)\equiv a$. The conserved input power is
\begin{equation}\nonumber
P=\sqrt{\pi}aA^2.
\end{equation}
The validity of the Gaussian ansatz is supported by the quantitative comparison with the stationary soliton profile obtained by imaginary-time propagation, shown in the inset of Fig.~\ref{effective_potential}(a). 
{We note that this variational reduction describes the slowly varying solitonic collective coordinates and does not explicitly include the fast radiation component}, which will be examined separately later.

Substituting the ansatz in Eq.~(\ref{eq:gaussian_ansatz}) into Eq.~(\ref{NNLS}), we introduce the Lagrangian density
\begin{equation}
\begin{aligned}
\mathcal{L}
=&\frac{i}{2}\left(u^\ast\frac{\partial u}{\partial z}-u\frac{\partial u^\ast}{\partial z}\right)
-\frac{1}{2}\left|\frac{\partial u}{\partial x}\right|^2  
+\frac{\gamma}{2}|u|^4 \\
&+\frac{|u|^2}{2}\int_{-\infty}^{+\infty}R(x-\xi)|u(\xi)|^2d\xi
-\alpha^2 x^2 |u|^2 .
\end{aligned}
\end{equation}
Integrating over $x$, we obtain the averaged Lagrangian,
\begin{equation}\nonumber
L=\int_{-\infty}^{+\infty}\mathcal{L}dx.
\end{equation}
Applying the Euler-Lagrange equations \cite{Borisreview,KIVSHAR1995353,perez1996low} yields the following Ermakov-like equation for the soliton width,
\begin{equation}
\ddot{a}=\frac{1}{a^3}-2\alpha^2 a
-\frac{P\gamma}{\sqrt{2\pi}a^2}
-\frac{2Pa}{\sqrt{\pi}(2a^2+\sigma^2)^{3/2}}
\equiv -\frac{\partial V}{\partial a},
\label{ermakov}
\end{equation}
where the dot denotes differentiation with respect to $z$.

%The two profiles are in excellent agreement, with a fidelity of $0.997$

Eq.~(\ref{ermakov}) also provides a natural starting point for defining the adiabatic reference trajectory used below. In the adiabatic limit, the control parameters vary sufficiently slowly that the soliton width remains close to the instantaneous equilibrium position of the effective potential. Denoting this reference width by $a_c(z)$, one has
\begin{equation}
2\alpha^2 a_c-\frac{1}{a_c^3}
+\frac{P\gamma}{\sqrt{2\pi}a_c^2}
+\frac{2Pa_c}{\sqrt{\pi}(2a_c^2+\sigma^2)^{3/2}}=0,
\label{adiabatic_condition}
\end{equation}
which determines the minimum of the effective potential $V\equiv V(a)$ at each propagation distance $z$. As illustrated in Fig.~\ref{effective_potential}(a), changing the characteristic response length $\sigma$ shifts the shape and minimum position of the effective potential, thereby modifying the stationary soliton width. In particular, a smaller $\sigma$ strengthens localization and moves the potential minimum toward smaller $a$, corresponding to a narrower stable soliton.

To quantify the  adiabaticity condition, we write the soliton width as
\begin{equation}
a(z)=a_c(z)+\delta a(z),
\end{equation}
where $\delta a(z)$ denotes a small deviation from the instantaneous equilibrium trajectory. Substituting this expression into Eq.~(\ref{ermakov}) and linearizing around $a_c$ \cite{berry2007dynamics, Hanpra206}, we obtain
\begin{equation}
\delta \ddot a+\Omega^2(z)\delta a \simeq -\ddot a_c,
\label{linearized}
\end{equation}
where
\begin{equation}
\Omega^2(z)=
2\alpha^2+\frac{3}{a_c^4}
-\frac{2P\gamma}{\sqrt{2\pi}a_c^3}
+\frac{2P}{\sqrt{\pi}}
\frac{\sigma^2-4a_c^2}{(2a_c^2+\sigma^2)^{5/2}}
\label{Omega2}
\end{equation}
is the squared frequency of small width oscillations around the instantaneous equilibrium. Eq.~(\ref{linearized}) shows that adiabatic following requires the motion of the equilibrium point to be sufficiently slow compared with the intrinsic response of the width mode. A convenient adiabaticity condition is therefore
\begin{equation}
\eta(z)=\frac{|\ddot a_c|}{\Omega^2 a_c}\ll 1.
\label{adiabaticity}
\end{equation}
This relation makes the role of nonlocality explicit: unlike in the local case, the adiabatic condition depends directly on the characteristic response length $\sigma$, which modifies the curvature of the effective potential and hence the intrinsic response timescale of the soliton width. We also note that the adiabatic reference and the corresponding adiabaticity condition depend not only on the control parameters $\sigma$, $\gamma$, and $\alpha$, but also on the input power $P$, which sets the overall strength of the nonlinear self-action and thus modifies both the equilibrium width and the local curvature of the effective potential.

To connect this criterion with the actual shortcut protocol, we prescribe a smooth compression trajectory $a(z)$ and use Eq.~(\ref{ermakov}) to reconstruct the corresponding $\sigma(z)$. The associated reference width $a_c(z)$ is then obtained from the equilibrium condition in Eq.~(\ref{adiabatic_condition}). Numerical analysis with the parameters used in this work shows that the adiabaticity criterion in Eq.~(\ref{adiabaticity}), expressed as $\max_z\eta(z)\lesssim 10^{-3}$, requires $z_f\gtrsim 48.18$. Accordingly, for $z_f=50$, the prescribed trajectory $a(z)$, shown by the orange solid line in Fig.~\ref{effective_potential}(b), is nearly indistinguishable from the reference width $a_c(z)$, shown by the blue dash-dotted line, indicating that the system remains in the adiabatic regime. By contrast, for $z_f=3$, one finds $\max_z\eta(z)>5$, so the evolution is manifestly nonadiabatic. In this case, the corresponding reference width, plotted as the green dashed line in Fig.~\ref{effective_potential}(b), deviates significantly from the prescribed trajectory in the intermediate region, even though the boundary values still coincide.

This distinction between $a(z)$ and $a_c(z)$ provides the physical origin of the fidelity loss in strongly accelerated protocols. When the protocol duration is long, the prescribed width follows the slowly moving equilibrium and the system remains close to the instantaneous stationary soliton, leading to high final fidelity. When $z_f$ is reduced, however, the required modulation of $\sigma(z)$ becomes steeper, the equilibrium point $a_c(z)$ moves more rapidly, and the condition $\eta(z)\ll1$ is violated. As a result, the width dynamics is no longer able to adiabatically follow the effective-potential minimum, nonadiabatic excitations are generated, and the final profile deviates more strongly from the target stationary soliton.

The influence of nonlocality can be further clarified by considering the two limiting regimes determined by the ratio between the response length $\sigma$ and the soliton width $a$. In the weakly nonlocal regime, $\sigma\ll a$, the nonlocal contribution in Eq.~(\ref{ermakov}) can be expanded as
\begin{equation}
-\frac{2Pa}{\sqrt{\pi}(2a^2+\sigma^2)^{3/2}}
\approx
-\frac{P}{\sqrt{2\pi}a^2}
+\frac{3P\sigma^2}{4\sqrt{2\pi}a^4}.
\label{weak_nonlocal}
\end{equation}
To leading order, the nonlocal term has the same $a$ dependence as the local Kerr contribution. This indicates that weak nonlocality mainly acts as a renormalization of the effective local nonlinearity, while genuine nonlocal corrections only appear at higher order through the $a^{-4}$ term.

By contrast, in the strongly nonlocal regime, $\sigma\gg a$, one finds
\begin{equation}
-\frac{2Pa}{\sqrt{\pi}(2a^2+\sigma^2)^{3/2}}
\approx
-\frac{2P}{\sqrt{\pi}\sigma^3}a
+\frac{6P}{\sqrt{\pi}\sigma^5}a^3.
\label{strong_nonlocal}
\end{equation}
In this limit, the leading nonlocal contribution becomes linear in $a$, which is formally equivalent to an additional harmonic restoring term in the effective width dynamics. This shows that strong nonlocality reshapes the nonlinear self-action into an effective confinement with strength directly controlled by $\sigma^{-3}$. As a result, modulation of the characteristic response length provides a particularly transparent and physically natural control knob for constructing shortcut protocols.

These two limiting cases also clarify the adiabaticity condition in Eq.~(\ref{adiabaticity}). In the weakly nonlocal regime, the dynamics remains close to that of a local nonlinear medium, with nonlocality providing only perturbative corrections. In the strongly nonlocal regime, however, the response length $\sigma$ directly controls the effective curvature of the width potential and thus the intrinsic oscillation frequency $\Omega$, making the adiabatic following and its breakdown especially sensitive to the modulation of the nonlocal response. This observation is useful below when comparing different STA protocols. In both limits, the input power $P$ controls the overall strength of the nonlocal contribution: it acts as an effective renormalization of self-focusing in the weakly nonlocal regime, while in the strongly nonlocal regime it sets the magnitude of the effective harmonic-like confinement. In the following, the input power $P$ is kept fixed and serves as a system parameter rather than a control variable. Nevertheless, it plays an important role in shaping the effective potential landscape and therefore influences both the adiabatic reference trajectory and the strength of the nonlocal contribution in the weakly and strongly nonlocal regimes.

\section{Inverse Engineering}
\label{section3}

Rapid soliton compression in nonlocal media can be implemented by modulating different physical parameters, including the characteristic length of the nonlocal response, the Kerr nonlinearity, and the external trapping strength. In what follows, we construct STA protocols based on the reduced width dynamics in Eq.~(\ref{ermakov}) and compare the resulting control smoothness and fidelity performance for these three control knobs.

\subsection{Control via the characteristic response length}
\label{1}

The characteristic response length $\sigma$ determines the degree of spatial averaging in a nonlocal medium and therefore strongly influences the spatial confinement of optical or matter-wave packets. Increasing $\sigma$ enhances the averaging effect, typically leading to broader solitons with lower peak intensity, whereas decreasing $\sigma$ strengthens localization and results in narrower beam profiles \cite{bang2002collapse}. Modulating $\sigma$ therefore provides a direct and physically transparent way to tailor the soliton width and its propagation dynamics. {In the adiabatic case, the characteristic length is taken to vary linearly as $\sigma(z)=\sigma_0+\beta z$, leading to a slow monotonic compression \cite{liang2021adiabatic}. This linear variation is feasible in NLCs, where the nonlocal response length can be tuned by the applied bias voltage \cite{hu2006nonlocality}.}

Using the reduced width equation~(\ref{ermakov}), the adiabatic reference trajectory is defined by the instantaneous equilibrium width $a_c(z)$ satisfying Eq.~(\ref{adiabatic_condition}). To ensure a smooth interpolation between the boundary values, we adopt a minimum-jerk trajectory for the soliton width,
\begin{eqnarray}
a(z)=a_i+(a_f-a_i)\left(10s^3-15s^4+6s^5\right),
\label{eq:poly_ansatz}   
\end{eqnarray}
with $s=z/z_f$. This form automatically satisfies
\begin{equation}
\begin{aligned}
a(0) &= a_c(0)=a_i, &\qquad a(z_f) &= a_c(z_f)=a_f, \\
\dot a(0) &= \dot a(z_f)=0, &\qquad \ddot a(0) &= \ddot a(z_f)=0.
\end{aligned}
\end{equation}
For the parameters $P=20$, $\gamma=0.1$, $\alpha=0.1$, $\sigma(0)\equiv \sigma_0=5$, and $\sigma (z_f)\equiv \sigma_f=0.5$, Eq.~(\ref{adiabatic_condition}) yields $a_i=1.014$ and $a_f=0.316$.

As shown in Fig.~\ref{effective_potential}(b), both the adiabatic and STA protocols compress the soliton width from $a(0)$ to $a(z_f)$. The inverse-engineered trajectory $a(z)$ reaches the target width within a propagation distance $z_f=3$, whereas the adiabatic reference requires a much longer distance, $z_f=50$. 

\begin{figure}
    \centering
    \includegraphics[width=\linewidth]{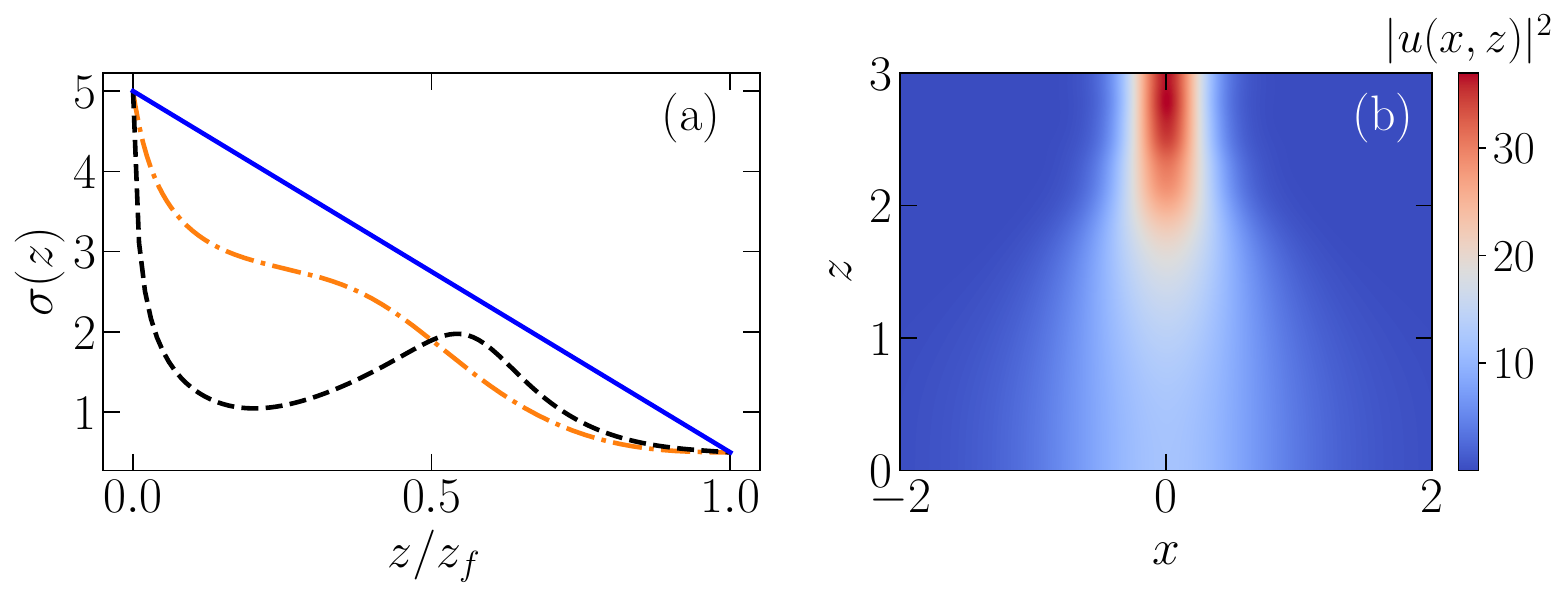}
    \caption{(a) Response length $\sigma(z)$ for the STA protocol with $z_f=3$ (orange dash-dotted) and $z_f=1$ (black dashed), together with the adiabatic protocol $\sigma(z)=\sigma_0+\beta z$ (blue solid) with $z_f=50$. (b) Spatiotemporal evolution of the soliton wave packet under the STA protocol when $z_f=3$. Other parameters are $\sigma_0=5$ and $\beta=-0.09$.}
    \label{F1}
\end{figure}

Eq.~(\ref{ermakov}) is then inverted to obtain the required $\sigma(z)$ in the STA protocol, as shown in Fig.~\ref{F1}(a). The orange dash-dotted curve corresponds to the protocol with $z_f=3$. In principle, the STA construction allows compression over very short propagation distances while maintaining high fidelity. However, when the protocol duration becomes too short, for example at $z_f=1$, the required variation of $\sigma(z)$ becomes much steeper, as indicated by the black dashed curve. Such rapid modulation may be difficult to implement experimentally and may induce additional distortions owing to the physical limitations of the medium \cite{rotschild2008incoherent, peccianti2002nonlocal}.

Fig.~\ref{F1}(b) shows the evolution of the soliton density distribution $|u(x,z)|^2$ obtained from direct numerical simulations of the full NNLSE. The compression dynamics are substantially accelerated under the STA protocol, while the soliton profile remains well controlled throughout the evolution. The final profile obtained from the STA simulation agrees closely with the stationary profile computed by imaginary-time propagation, confirming the validity of the protocol. To quantify the accuracy of the accelerated compression, we define the fidelity as
\begin{equation}
F=\left|\langle \tilde{u}_f(x)\,|\,u(x,z_f)\rangle\right|^2,
\end{equation}
where $u(x,z_f)$ is the final field produced by the STA protocol, and $\tilde{u}_f(x)$ is the target stationary profile obtained by imaginary-time propagation. The initial state used in the STA evolution is also prepared from the corresponding stationary soliton calculated by imaginary-time propagation. As already shown in the inset of Fig.~\ref{effective_potential}(a), this stationary profile is very close to, although not exactly identical with, the Gaussian ansatz. For the protocol shown in Fig.~\ref{F1}, the fidelity reaches $F=0.9993$, demonstrating that high-fidelity soliton compression can be achieved by inverse engineering of the characteristic response length.

\begin{figure}
    \centering
    \includegraphics[width=0.9\linewidth]{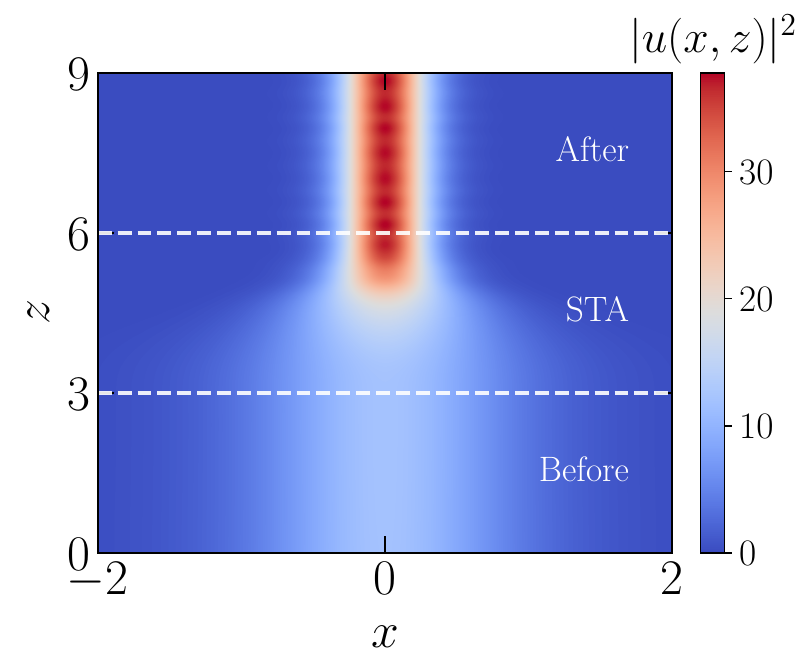}
    \caption{Complete spatiotemporal evolution of the soliton under the $\sigma(z)$-based STA protocol with $z_f = 3$. White dashed lines separate three stages: stable propagation of the initial broad soliton $\sigma_0 = 5$ (Before), rapid compression under the inverse-engineered $\sigma(z)$ (STA), and stable propagation of the compressed soliton at constant $\sigma_f$ (After). Other parameters are the same as those in Fig.~\ref{F1}.}
    \label{fig:complete_evolution}
\end{figure}

{To further illustrate the full implementation process of the STA protocol, Fig.~\ref{fig:complete_evolution} shows the complete propagation dynamics of the representative $\sigma(z)$-based strategy with $z_f=3$. 
The evolution consists of three consecutive stages: stable propagation of the initial broad soliton at fixed $\sigma_0=5$, rapid compression under the inverse-engineered $\sigma(z)$, and stable propagation of the compressed narrow soliton at fixed $\sigma_f=0.5$. This before-during-after evolution demonstrates that the STA protocol connects the initial and final stationary soliton states within a shortened propagation distance, while avoiding appreciable residual excitation.}

\subsection{Control via the Kerr nonlinearity}

An increase in the Kerr coefficient $\gamma$ enhances the nonlinear refractive-index response and strengthens self-phase modulation. As a result, the nonlinear phase accumulation becomes larger and the field energy is more strongly concentrated around the soliton center. This shifts the balance between diffraction and nonlinearity toward tighter localization, thereby reducing the soliton width and increasing the peak amplitude \cite{khater2025dynamic, zhang2025nonlinear}.

\begin{figure}
    \centering
    \includegraphics[width=\linewidth]{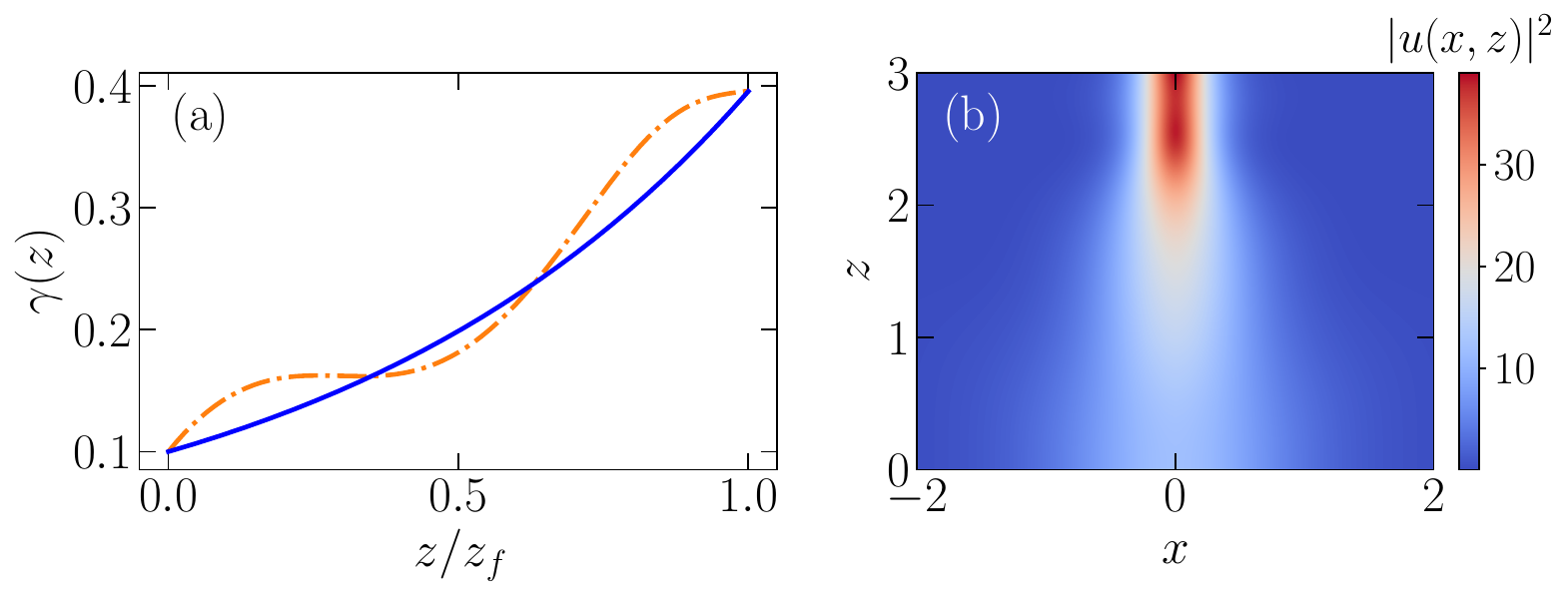}
    \caption{(a) Comparison between the inverse-engineered Kerr nonlinearity $\gamma(z)$ for $z_f=3$ (orange dash-dotted) and the adiabatic protocol $\gamma(z)=0.1e^{\gamma_0 z}$ with $\gamma_0=0.0275$ and $z_f=50$ (blue solid). (b) Spatiotemporal evolution of the soliton wave packet under the STA protocol. Other parameters are the same as those in Fig.~\ref{effective_potential}.}
    \label{F2}
\end{figure}

We next consider soliton compression in a nonlocal medium with fixed response length $\sigma$ and fixed waveguide strength $\alpha^2$. For the adiabatic reference, we choose an exponential nonlinearity profile,
\begin{equation}
\gamma(z)=0.1e^{\gamma_0 z},
\end{equation}
where $\gamma_0$ is a constant.{Exponentially varying nonlinearity profiles have been analyzed in fibers and amplifiers with distributed parameters and implemented via longitudinal engineering in tapered and photonic crystal fibers \cite{kruglov2003exact, sharafali2021self}.}
Using the adiabatic reference and the minimum-jerk interpolation in Eq.~(\ref{eq:poly_ansatz}), one can invert Eq.~(\ref{ermakov}) to determine the required $\gamma(z)$ for the STA protocol. Imposing the boundary values $\gamma(0)=0.1$ and $\gamma(z_f)=0.396$ leads to the same target compression as that obtained by modulating $\sigma$.

Fig.~\ref{F2}(a) compares the inverse-engineered STA profile of $\gamma(z)$ with the corresponding adiabatic protocol. The STA protocol reaches the target compressed width within a much shorter propagation distance, while the adiabatic protocol requires a long and gradual increase of the nonlinearity. Fig.~\ref{F2}(b) shows the corresponding spatiotemporal evolution of the soliton wave packet. In this case, the Kerr nonlinearity acts as a direct compression knob by strengthening the local self-focusing contribution during propagation.

\subsection{Control via the waveguide strength}

\begin{figure}
    \centering
    \includegraphics[width=0.51\textwidth]{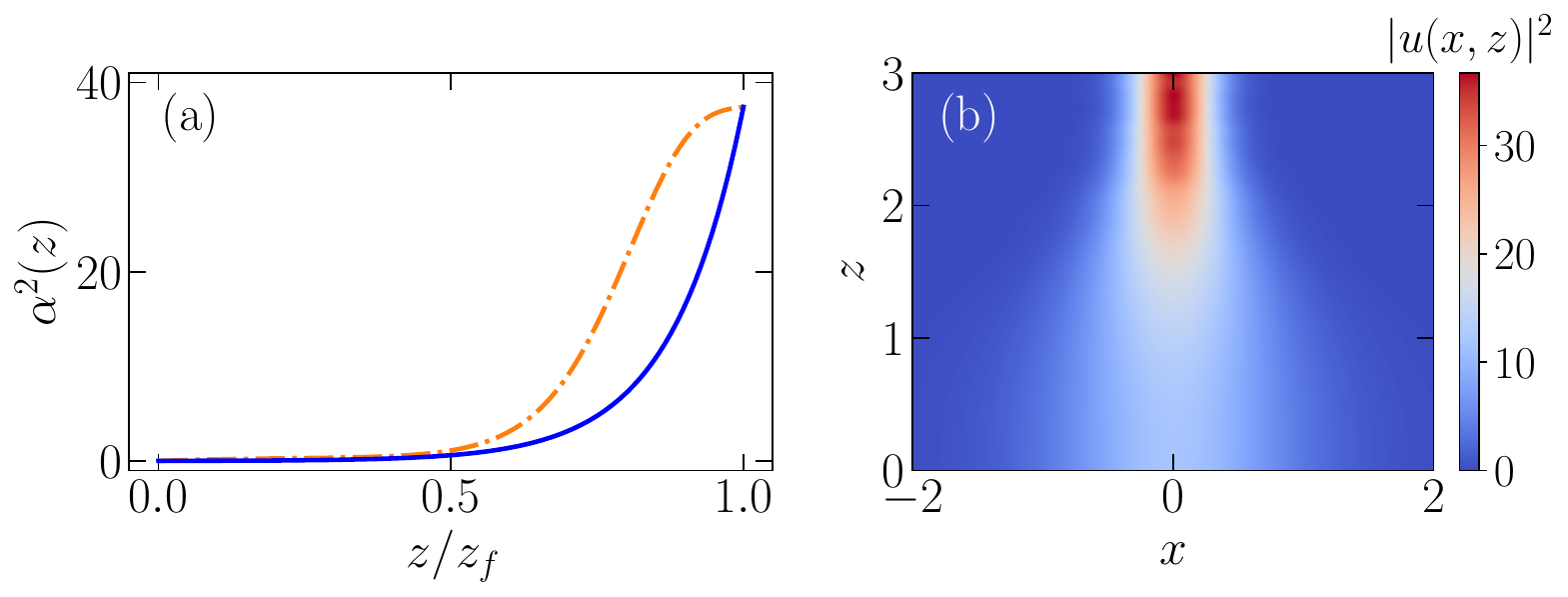}
    \caption{(a) Waveguide strength $\alpha^2(z)$ designed through inverse engineering for the STA protocol with $z_f=3$ (orange dashed), together with the adiabatic protocol $\alpha^2(z)=0.01e^{\alpha_0 z}$ with $\alpha_0=0.165$  and $z_f=50$ (blue solid). (b) Spatiotemporal evolution of the soliton wave packet under the STA protocol. Other parameters are the same as in Fig.~\ref{effective_potential}.}
    \label{F3}
\end{figure}

We finally consider modulation of the external waveguide strength. The parabolic potential in Eq.~(\ref{parabolic}) provides transverse confinement and corresponds, in optical language, to a quadratic refractive-index profile. Increasing $\alpha^2$ enhances the curvature of this profile, leading to tighter confinement and improved beam localization \cite{kong2016spatiotemporal, koutsokostas2024particle, belyaeva2021interaction}.

To enable direct comparison with the previous protocols, we again construct the STA trajectory from the same reduced width dynamics and impose the same boundary conditions on $a(z)$. In the adiabatic protocol, the waveguide strength is taken to vary exponentially,
\begin{equation}
\alpha^2(z)=0.01e^{\alpha_0 z},
\end{equation}
with constant $\alpha_0$.  {Such exponential variation of the waveguide strength can be realized in longitudinally tapered graded-index waveguides \cite{triki2019propagation}.} Once the minimum-jerk width trajectory in Eq.~(\ref{eq:poly_ansatz}) is fixed, Eq.~(\ref{ermakov}) can be inverted to determine the corresponding $\alpha^2(z)$. By imposing the boundary values $\alpha^2(0)=0.01$ and $\alpha^2(z_f)=37.41$, one obtains rapid compression of the soliton through propagation-dependent external confinement.

Fig.~\ref{F3}(a) shows the inverse-engineered STA profile of the waveguide strength together with the adiabatic reference. As in the previous cases, the STA protocol strongly reduces the propagation distance required to reach the target width. However, for very short distances, e.g. $z_f=0.6$, the required $\alpha^2(z)$ may become negative (not shown), corresponding to a locally antiguiding parabolic potential. In that regime, the refractive index has a minimum at the center, which tends to expel the optical beam rather than confine it \cite{dovgiy2015discrete}. This indicates an important practical limitation of the waveguide-based protocol. Fig.~\ref{F3}(b) displays the corresponding spatiotemporal evolution of the field intensity $|u(x,z)|^2$, showing that substantial compression can nevertheless be achieved within a finite propagation distance.

\section{Discussion}
\label{sec4}

\begin{figure}
    \centering
    \includegraphics[width=\linewidth]{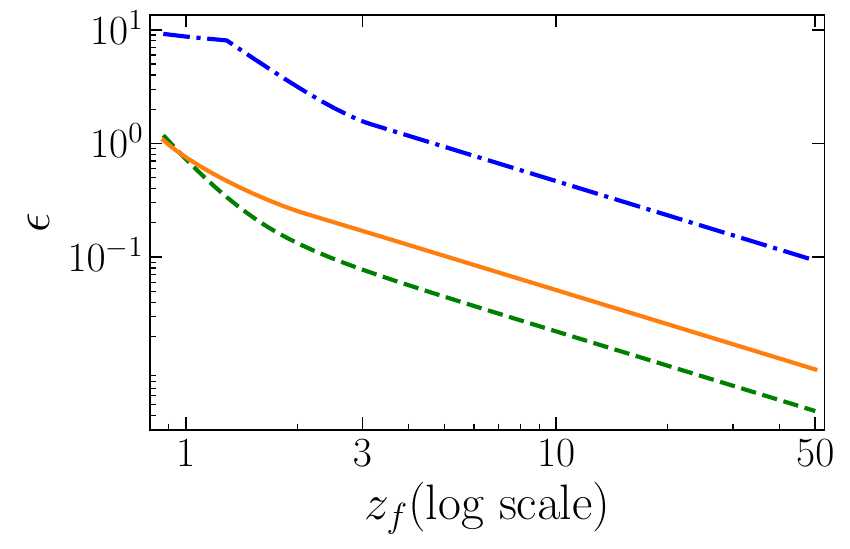}
    \caption{Normalized total variation of the inverse-engineered control parameters. The green dashed, blue dash-dotted, and orange solid curves correspond to $\alpha^2(z)$, $\gamma(z)$, and $\sigma(z)$, respectively. The parameters are the same as those in Figs.~\ref{F1}-\ref{F3}.}
    \label{fig:variation}
\end{figure}

We have achieved rapid compression of the soliton width from $a_i$ to $a_f$ by inverse engineering three different control parameters: the characteristic response length $\sigma(z)$, the Kerr nonlinearity $\gamma(z)$, and the transverse confinement strength $\alpha^2(z)$. To compare these strategies under the same compression target, we fix the input power and examine how the required control modulation depends on the final propagation distance $z_f$.

To quantify the smoothness of the control protocols, we define the relative normalized total variation as
\begin{equation}
\epsilon=\frac{1}{z_f\left[\rho_{\max}-\rho_{\min}\right]}
\sum_i \left|\frac{\rho_{i+1}-\rho_i}{\rho_i}\right|,
\qquad \rho\in\{\sigma,\gamma,\alpha^2\},
\end{equation}
where $\rho_{\max}$ and $\rho_{\min}$ denote the extremal values of the control parameter over $z\in[0,z_f]$, and $\rho_i$ are values at adjacent grid points. This quantity measures the cumulative relative variation of the control parameter, normalized by both its overall amplitude and the protocol length. A larger $\epsilon$ corresponds to stronger and faster parameter modulation, implying a less smooth and potentially more demanding protocol. Conversely, a smaller $\epsilon$ indicates a smoother control profile and improved experimental feasibility.

\begin{figure}
    \centering
    \includegraphics[width=\linewidth]{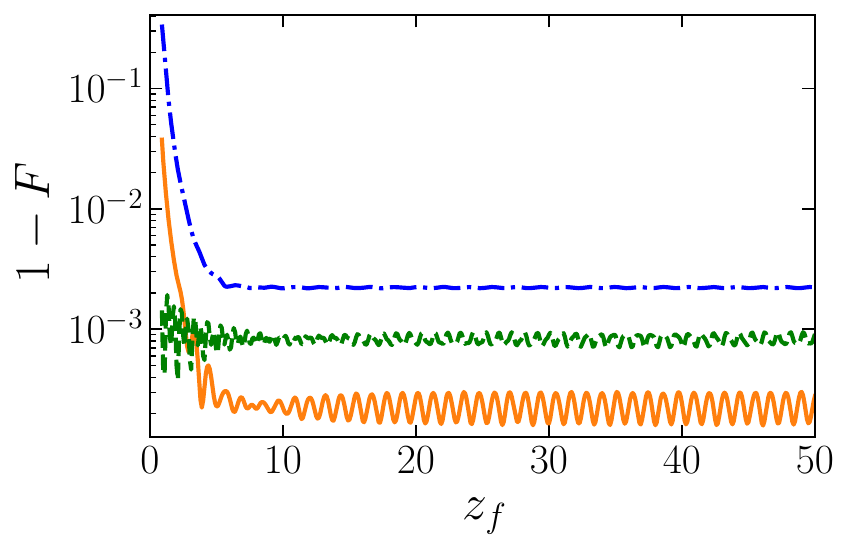}
    \caption{Final infidelity $1-F$ as a function of the propagation distance $z_f$ for the three STA protocols. The green dashed, blue dash-dotted, and orange solid curves correspond to $\alpha^2(z)$, $\gamma(z)$, and $\sigma(z)$, respectively. The parameters are the same as those in Figs.~\ref{F1}-\ref{F3}.}
    \label{fig:infidelity}
\end{figure}

Fig.~\ref{fig:variation} shows the normalized total variation of the three control parameters as a function of $z_f$. In the short-distance regime, the $\sigma(z)$-based protocol becomes particularly favorable in the extremely short-distance limit around $z_f\sim 1$, where it yields the smallest variation among the three strategies. This reflects the fact that, in a nonlocal medium, modulation of $\sigma$ reshapes the effective restoring landscape in a comparatively smooth manner. The protocol based on $\alpha^2(z)$ also exhibits relatively small variation over a broader range of propagation distances, indicating that external harmonic confinement can efficiently assist the compression once the protocol is not pushed to the extreme nonadiabatic limit. However, for sufficiently short distances, the inverse-engineered $\alpha^2(z)$ may even become negative, corresponding to an antiguiding regime and thus imposing an additional practical limitation. By contrast, the protocol based on $\gamma(z)$ displays the largest variation and the strongest sensitivity to $z_f$, indicating that direct modulation of the local Kerr self-focusing generally requires the most aggressive control.

The monotonic decrease of $\epsilon$ with increasing $z_f$ reflects the approach to the adiabatic limit, where longer propagation distances reduce the amount of excitation generated during the compression and therefore relax the required strength of the control modulation. In particular, the confinement-based protocol becomes less demanding once the available propagation distance is sufficiently long. However, control smoothness alone does not fully characterize the performance of the STA protocols. To assess the dynamical accuracy of the compression, we also examine the final infidelity as a function of $z_f$.

Fig.~\ref{fig:infidelity} presents the corresponding final infidelity $1-F$. Comparison between Figs.~\ref{fig:variation} and \ref{fig:infidelity} shows that control smoothness and final fidelity characterize distinct aspects of STA performance. In particular, in the very short-distance regime, for example at $z_f=1$ as shown in Fig.~\ref{F1}(a), a smoother control profile does not necessarily yield a lower infidelity, revealing a trade-off between implementation smoothness and target-profile accuracy. 
% \textcolor{red}{More broadly, the infidelity $1 - F$ reflects the impact of radiation emission and non-Gaussian deformations, which are not captured within the variational approximation, on soliton compression. The high fidelity achieved indicates that these effects remain negligible during propagation, whereas the increase of $1 - F$ with decreasing $z_f$ reveals the nonadiabatic excitations, representing the cost of the shortcut.}

This behavior can be understood from the structure of the reduced width dynamics in the weakly and strongly nonlocal limits. In the weakly nonlocal regime ($\sigma \ll a$), the nonlocal contribution mainly renormalizes the effective nonlinear interaction, so that modulation of $\sigma$ introduces only a mild correction to the width dynamics. In the strongly nonlocal regime ($\sigma \gg a$), the nonlocal term becomes approximately linear in $a$, and the dynamics reduces to an effective harmonic-like restoring force whose strength is directly controlled by $\sigma$. In this regime, tuning $\sigma$ effectively reshapes the effective potential landscape, leading to smoother control trajectories. This explains why the $\sigma(z)$-based protocol tends to exhibit reduced parameter variation compared with $\gamma(z)$ or $\alpha^2(z)$, which act more locally on the nonlinear or external contributions. However, the final infidelity depends on the full nonlinear evolution and cannot be inferred from control smoothness alone. As a result, in the very short-distance regime, different control strategies may perform differently depending on the chosen criterion, reflecting the intrinsic trade-off between smoothness and dynamical accuracy.

\begin{figure}
    \centering
    \includegraphics[width=\linewidth]{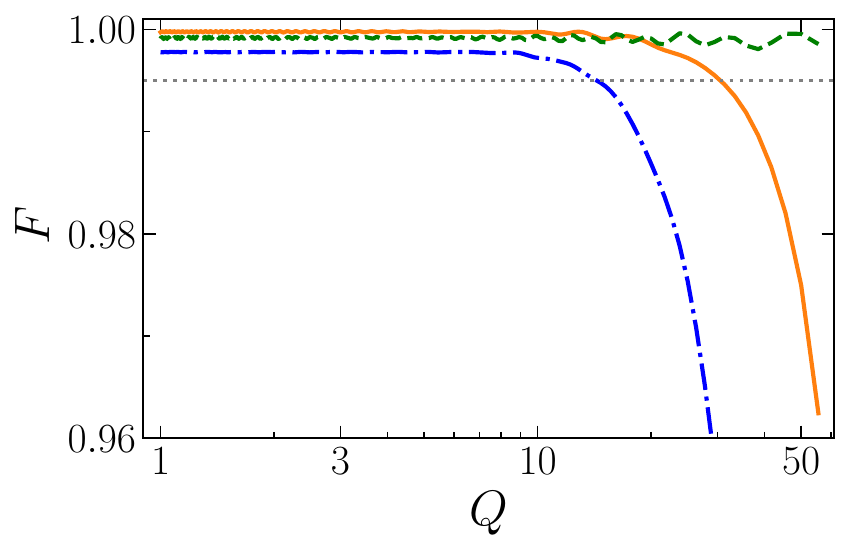}
    \caption{Fidelity $F$ as a function of the acceleration factor $Q = z_{\mathrm{ad}}/z_f$ for the three STA protocols, where $z_{\mathrm{ad}} = 50$ is the adiabatic reference distance. The orange solid, blue dash-dotted, and green dashed curves correspond to $\sigma(z)$, $\gamma(z)$, and $\alpha^2(z)$, respectively. The horizontal dotted line marks the high-fidelity threshold $F = 0.995$. The parameters are the same as those in Figs.~\ref{F1}--\ref{F3}.}
    \label{efficiency}
\end{figure}

{To compare shortcut efficiency and final field accuracy, we introduce the acceleration factor
\begin{equation}
Q=\frac{z_{\rm ad}}{z_f},
\end{equation}
where $z_{\rm ad}=50$ is the adiabatic reference distance. A larger $Q$ therefore indicates a shorter and more efficient shortcut. The comparison in Fig.~\ref{efficiency} shows that, under the constraint $F\geq 0.995$, the $\sigma$-, $\gamma$-, and $\alpha^2$-based protocols reach $z_f=1.7$, $3.5$, and $0.9$, corresponding to $Q=29.4$, $14.3$, and $55.6$, respectively. The $\alpha^2$-based protocol therefore enables the strongest acceleration at high fidelity, although in this regime the inverse-engineered $\alpha^2(z)$ may approach the antiguiding limit, imposing a practical bound on further acceleration. The optimal $z_f$ should thus be chosen by balancing acceleration against fidelity and implementation constraints.}

Taken together, these results show that no single control protocol is uniformly superior across all propagation distances when both smoothness and fidelity are taken into account. In particular, modulation of the characteristic response length $\sigma(z)$ becomes especially favorable from the viewpoint of control smoothness in the extremely short-distance regime, whereas the $\alpha^2(z)$-based protocol can provide the largest acceleration under a high-fidelity constraint, subject to practical limitations associated with possible antiguiding behavior. Thus, the choice of control knob should depend on the desired balance among implementation smoothness, final fidelity, shortcut efficiency, and experimental feasibility.

% \textcolor{red}{To quantify the performance of each protocol more systematically, we define a protocol efficiency
% \begin{equation}
% Q(z_f) = \frac{z_{\mathrm{ad}}}{z_f} F(z_f),
% \label{eq:efficiency}
% \end{equation}
% where $z_{\mathrm{ad}}=50$ is the adiabatic reference distance. For instance, the $\alpha^2$-based protocol reaches its maximum efficiency at $z_f = 0.9$ with $F = 0.998$. For the $\sigma$- and $\gamma$-based protocols, the maximum efficiencies also occur at $z_f = 0.9$ and $z_f = 1.0$, respectively, but the corresponding fidelities are only $0.962$ and $0.746$. Therefore, we further impose a fidelity constraint $F \geq 0.995$, the $\sigma$- and $\gamma$-based protocols reach their optimal propagation distances at $z_f = 1.7$ and $z_f = 3.5$, respectively. These results indicate that shorter and more efficient shortcuts are in principle achievable, but the optimal choice of $z_f$ should balance the acceleration factor and the required fidelity.}

{Finally, we emphasize that variational approximation describes the slow collective motion of the soliton core and does not explicitly include fast linear waves that may be emitted during rapid compression. To quantify such radiation leakage beyond the variational description, we decompose the numerical field at each propagation distance as
\begin{equation}
u(x,z)=c(z)u_0(x,z)+\delta u(x,z),
\label{eq:bdg_decomp}
\end{equation}
where $u_0(x,z)$ is the instantaneous stationary soliton corresponding to the $\sigma(z)$-based protocol, and $c(z)=\langle u_0|u\rangle/\langle u_0|u_0\rangle$ is the projection coefficient. 
The residual field $\delta u$ contains both localized internal deformations of the soliton core and delocalized linear-wave radiation. As detailed in Appendix~\ref{app A}, we linearize the NNLSE around $u_0$ and use the instantaneous Bogoliubov-de Gennes modes to separate these two contributions. The delocalized part is identified as the fast-radiation component $\delta u_f$, whose fractional power $P_f(z)/P$ provides a diagnostic of the linear-wave leakage induced by the shortcut protocol.}

\begin{figure}
    \centering
    \includegraphics[width=\linewidth]{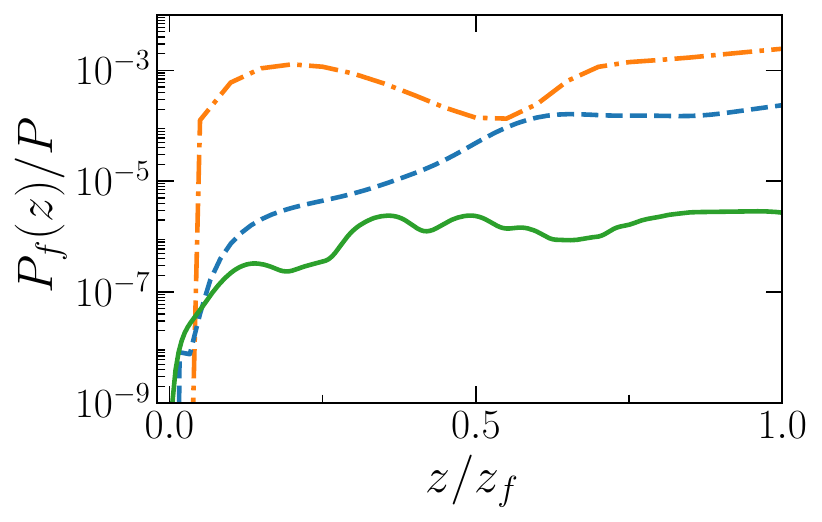}
    \caption{Fractional fast-radiation power $P_f(z)/P$ obtained from the linearized radiation analysis for the $\sigma(z)$-based STA compression protocol. The orange dash-dotted, blue dashed, and green solid curves correspond to \(z_f=1\), \(z_f=3\), and \(z_f=10\), respectively. The parameters are the same as those in Fig.~\ref{F1}.}
    \label{bdg}
\end{figure}

{The normalized fast-radiation power $P_f(z)/P$ is shown in Fig.~\ref{bdg}. As $z_f$ is reduced, $P_f(z)/P$ increases, indicating that stronger acceleration drives the soliton further away from adiabatic following and excites a larger linear-wave component. Nevertheless, even for the most aggressive case considered here, $z_f=1$, the radiation remains at the $10^{-3}$ level. For the representative protocol with $z_f=3$, it is further suppressed to the $10^{-5}$, consistent with the high fidelity $F=0.9993$ reported before. This diagnostic shows that the emitted linear-wave component remains negligible compared with the total optical power in the parameter regime considered here, supporting the consistency of the variational approximation for designing STA protocol.}

\section{Conclusion and outlook}
\label{sec:5}

In conclusion, we have developed an inverse-engineered shortcut-to-adiabaticity framework for the rapid compression of optical solitons in nonlocal media. Starting from a generalized NNLSE, we employed a Gaussian variational ansatz to derive an effective Ermakov-like equation for the soliton width, which provides a reduced dynamical description for constructing finite-distance STA protocols. Within this framework, we investigated three control strategies based on modulation of the characteristic nonlocal length, the Kerr nonlinearity, and the external parabolic confinement. Our results show that all three protocols can realize high-fidelity soliton compression over strongly reduced propagation distances compared with the corresponding adiabatic evolution.

A systematic comparison of control smoothness, final fidelity, and shortcut efficiency reveals that the three protocols exhibit different advantages. Modulation of the characteristic nonlocal length provides a comparatively smooth control profile and remains highly competitive in fidelity, especially in the short-distance regime. The confinement-based protocol can reach the largest acceleration under a high-fidelity threshold, although its practical implementation may be limited by the possible onset of an antiguiding regime at very short distances. These results show that control smoothness, final fidelity, and acceleration efficiency characterize distinct aspects of STA performance, and that no single protocol is uniformly preferred when different performance criteria and implementation constraints are taken into account.

We have also examined the physical implementation and robustness of the shortcut process. The complete before-during-after propagation dynamics confirms that the STA protocol connects the initial broad soliton and the final compressed soliton while maintaining stable propagation outside the shortcut stage. In addition, the instantaneous Bogoliubov-de Gennes radiation diagnostic shows that the emitted linear-wave component remains weak in the parameter regime considered, reaching only the $10^{-3}$ level even for the strongest acceleration considered. These observations indicate that radiation leakage does not significantly affect the high-fidelity compression and support the variational approximation method for designing STA.

From a physical perspective, our analysis further clarifies the role of nonlocality in fast soliton control. In the weakly nonlocal regime, the nonlocal response mainly acts as a perturbative correction to effective local self-focusing, whereas in the strongly nonlocal regime it gives rise to an effective harmonic-like restoring contribution in the width dynamics. As a result, the characteristic response length emerges not merely as a material parameter, but as a genuine dynamical resource for shortcut design. The discussion of experimentally relevant control parameters, including bias-voltage tuning in nematic liquid crystals, longitudinally engineered nonlinear media, and tapered graded-index waveguides, further connects the proposed protocols with feasible physical platforms.

The present work can be extended in several directions. On the theoretical side, it would be interesting to go beyond the Gaussian variational ansatz and explore more accurate reduced descriptions in regimes where non-Gaussian deformations become important \cite{kong2013dark,Shenpre}. It would also be worthwhile to investigate STA protocols combined with optimal control \cite{Tangyoupra} for multidimensional nonlocal solitons, vector solitons, and interacting multisoliton states. On the practical side, the present method may be applied to other nonlinear-wave platforms and photonic functionalities \cite{saha2023engineering, errando2019low}, including beam shaping, all-optical switching, and beam steering in experimentally accessible nonlocal media. More broadly, these results suggest that shortcut-based control in nonlocal systems provides a promising route toward fast and robust manipulation of self-trapped wave packets.

\section*{ACKNOWLEDGMENTS}
This work is partially supported from NSFC (12075145, 12474362 and 12174243), STCSM (2019SHZDZX01-ZX04),  the Basque Government through Grant No. IT1470-22, and PID2021-126273NB-I00.  X.C.  acknowledges the Severo Ochoa Centres of Excellence program through Grant No. CEX2024-001445-S and the CSIC–NSTC programme (BINST24012).

\appendix

\section{Linearized Radiation Analysis}
\label{app A}

{To separate localized internal deformations of the soliton from delocalized radiation waves, we linearize the NNLSE around the instantaneous stationary profile $u_0(x,z)$ used in the decomposition of Eq.~(\ref{eq:bdg_decomp}). Writing the perturbation as $\delta u(x,z)=p(x,z)+iq(x,z)$, with $p$ and $q$ real, we obtain the instantaneous Bogoliubov-de Gennes
(BdG) system
\begin{equation} 
\frac{\partial}{\partial z} \begin{pmatrix} p \\ q \end{pmatrix} = \begin{pmatrix} 0 & L_-(z) \\ - L_+(z) & 0 \end{pmatrix} \begin{pmatrix} p \\ q \end{pmatrix}, 
\end{equation}
where $L_\pm(z)$ are the linearized operators obtained from the NNLSE around $u_0(x,z)$ for the instantaneous value of the control parameter. The stationary profile satisfies the corresponding nonlinear stationary equation with propagation constant $\mu(z)$.}

{The eigenmodes of this instantaneous BdG-type problem can be classified into spatially localized modes, which describe internal deformations of the soliton core, and delocalized modes associated with linear-wave radiation. Among the localized modes, width-like breathing deformations are expected to be the dominant excitations captured by the variational description. In the numerical implementation, the localized modes, together with the neutral solitonic directions, are identified by their spatial localization around the soliton core and then orthonormalized into a basis $\chi_m(x,z)$. This basis defines the localized deformation subspace, onto which the residual field is projected as
$\mathcal{P}(z)\delta u=
\sum_m \langle \chi_m|\delta u\rangle \chi_m $.
The component of $\delta u$ outside this localized subspace is identified as the radiated field,
$\delta u_{f}
=
\delta u-\mathcal{P}(z)\delta u $,
with power
\begin{equation}
P_{f}(z)=\int |\delta u_{f}(x,z)|^2 dx .
\end{equation}
The normalized quantity $P_{f}(z)/P$ is used in Fig.~\ref{bdg} to estimate the fraction of the total power emitted into linear radiation during the inverse-engineered compression.}

\bibliography{nonlocal}

\end{document}